\documentclass{ws-procs10x7}

\newcolumntype{d}[1]{D{.}{.}{#1}}

\def\Journal#1#2#3#4{{\it #1} {\bf #2}, #3 (#4)}

\input symblibrary 

\newcommand{\ltsim}{\raisebox{-1ex}
{\mbox{$\mbox{}\stackrel{\textstyle <}{\sim}\mbox{}$}}}

\makeindex
\begin{document}

\title{ Latest~Results~on~Orbitally~Excited~Strange\\
        Bottom~Mesons~with~the~CDF~II~Detector}

\author{Igor V. Gorelov\\
        For the CDF Collaboration
        }

\address{Department of Physics and Astronomy, University~of~New~Mexico, Albuquerque, NM 87131, USA \\
         E-mail: gorelov@fnal.gov\\
         (\small talk given at ICHEP~2006,~July~26~-August~2,~2006,~Moscow,~Russia)}


\twocolumn[\maketitle\abstract{
  We present the latest results on the spectroscopy of orbitally
  excited strange bottom mesons from $\sim$~1~fb$^{-1}$ of CDF
  data. The measurements are performed with fully reconstructed $B$
  decays collected by the CDF~II detector at $\sqrt{s} = 1.96$~TeV in both
  the di-muon and the fully hadronic trigger paths.
}
]

\section{Introduction}
\begin{figure}[b]
\centerline{\psfig{file=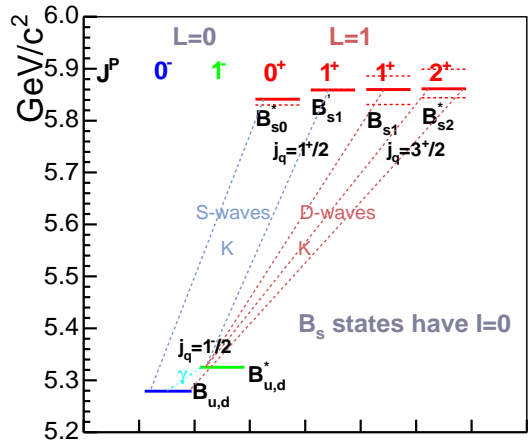,width=2.9in}}
\caption{ Spectroscopy of orbital \BsJ Mesons }
\label{fig:spectr}
\end{figure}

  Mesons containing one heavy quark are a useful laboratory to
  test QCD models.  In the limit of heavy quark mass $m_{Q}\to\infty$,
  heavy mesons' properties are governed by the dynamics of the light
  quark. As such, these states become ``hydrogen atoms'' of hadron
  physics. In this Heavy Quark Symmetry approach (see
  references~\cite{th:isgur1})
  the quantum numbers of the heavy and light quarks are separately
  conserved by the strong interaction. For the bottom \B-mesons the
  heavy bottom quark spin, $\mathbf{s_Q={\frac{1}{2}}^{+}}$, couples with the
  light anti-quark momentum \(\mathbf{j_q = s_q + L}\), where
  \( \mathbf{ s_q={\frac{1}{2}}^{-} } \) is the spin of the light anti-quark and
  $\mathbf{L}$ is its angular momentum. Hence for $P$-wave ($L=1$)
  mesons we obtain two \( j_q = {\frac{3}{2}}^{+} \) states, the 
  \(J^{P}=2^{+}{\rm ,}\,1^{+}\) states, and two 
  \( j_q = {\frac{1}{2}}^{+} \) states, the
  \(J^{P}=0^{+}{\rm ,}\,1^{+}\) states. 
  The sketch of the \BsJ- states with their possible strong decays 
  to lower lying non-strange  \(B_{u,d}\) mesons and \kaon- mesons are shown in
  Fig.~\ref{fig:spectr}. In our analysis we do not consider exotic
  modes (e.g.~\( \BsJ\to\Bs\piz \) ) for which isospin is not conserved.
  The mass predictions from several authors are summarized 
  in Table~\ref{tab:pred}.
\begin{table}
\tbl{Selected mass \( ({\rm in~\mevcc}) \)  predictions of \BsJ states \label{tab:pred}}
{\begin{tabular}{@{}lcccr@{}}
\toprule
Ref. & $B^{*}_{s0}$ & $B^{'}_{s1}$ & $B_{s1}$ & $B^{*}_{s2}$ \\ 
\colrule
\cite{th:godfrey} & 5830.0  & 5860.0 & 5860.0  & 5888.0 \\
\cite{mass:ebert} & 5841.0 & 5859.0 & 5831.0 & 5844.0 \\
\cite{mass:eichten} & - & - & 5834.0  & 5846.0 \\
\cite{mass:falk} & - & - & 5886.0 & 5899.0 \\
\botrule
\end{tabular}}
\end{table}
  The states $B^{*}_{s0}$ and $B^{'}_{s1}$ decaying with the \kaon-meson
  emitted in an $S$-wave are predicted to have a large width of 
  $(100 - 170)\mevcc$ (see e.g. reference~\cite{th:godfrey}). These states are
  experimentally difficult to identify. The states $B_{s1}$ and
  $B^{*}_{s2}$ decaying via $D$-waves should have a narrow width 
  predicted to be $(1 - 7)\mevcc$ for $B^{*}_{s2}$ and 
  $(<1.0 - 2.8)\mevcc$ for $B_{s1}$ 
  (see references~\cite{th:godfrey,mass:eichten,mass:falk}).  

  The experimental results on orbital \BsJ mesons are limited by
  observation of only one narrow state. The first evidence of the state was made
  by the OPAL Collaboration~\cite{exp:opal} and later confirmed by the DELPHI
  experiment~\cite{exp:delphi}. Recently the D\O ~Collaboration~\cite{exp:d0} 
  reported a narrow signal at the similar mass value as OPAL.

  In this report we present results on the search for and observation 
  of narrow \(J^{P}=1^{+}{\rm ,}\,2^{+}\) states of 
  strange bottom mesons.
\section{Triggers and Datasets}
  Our results are based on data collected with the \cdf2
  detector and corresponding to an integrated luminosity of
  $\sim1\invfb$. As $\pap$ collisions at 1.96\,TeV have an enormous
  inelastic total cross-section of \( \sim\,60\,\mbarn \), while \b-
  hadron events comprise only \( \approx\,20\,\mub\,\,(|\eta|<1.0) \),
  triggers selecting \b- hadron events are of vital importance. 
  The triggers set in the \cdf2 detector for \b- physics studies are 
  based on leptons and displaced tracks.

  One of these is a dimuon trigger with a low muon transverse momentum
  threshold of \( 1.5\gevc \).  It reconstructs at Level 1 track pairs
  in the CDF central tracker COT. The tracks are then matched to hits
  in the CDF muon chambers. At Level 3 a full reconstruction is made
  and a dimuon mass cut \( M(\mumu)\in[2.7,4.0]\gevcc \) around the
  mass of \jpsi is applied. This trigger saves \B-mesons through the
  mode \( \Bu\rightarrow\,\jpsi\Kp\,,\jpsi\rightarrow\mumu
  \)\footnote{Unless otherwise stated all references to the specific
  charge combination imply the charge conjugate combination as well.}.

  The other is the Two displaced Track Trigger.  It also reconstructs
  with the central tracker a pair of high \pt tracks at Level 1 and
  enables secondary vertex selection at Level 2 requiring each of
  these tracks to have impact parameter measured by the CDF silicon
  detector SVX II larger than 120\mum. The excellent impact parameter
  resolution of SVX II makes this challenging task possible. The Two
  Track Trigger is efficient for heavy quark hadron decay modes. It
  triggers another fundamental mode used in this analysis, namely \(
  \Bu\rightarrow\,\Dzb\pip\,,\Dzb\rightarrow\Kp\pim \).
\section{Event Selection}
  To avoid absolute mass scale systematic uncertainties, we search
  for narrow signatures in a mass difference distribution \( Q =
  M(\Bu\Km) - M(\Bu) - M(\Km) \), where M(\Bu\Km) is the invariant mass
  of the $\Bu\Km$ pair, M(\Bu) is the invariant mass of the
  $\Bu$ candidate, and M(\Km) is the $PDG$ mass of the kaon. In the mode
  (see Fig.~\ref{fig:spectr}) \( B^{*}_{s2}\rightarrow\,\Bu\Km \), the
  signal is expected to appear at \(70\mevcc\ltsim Q\ltsim 130\mevcc\).  If the
  mass of the state \( B^{*}_{s2} \) is above the \( \Bustar\Km \)
  threshold, an additional bump would emerge 45.78\mevcc below the other,
  due to the undetected \g from the radiative decay
  \(\Bustar\rightarrow\,\Bu\g\). The same is true for \(
  B_{s1}\rightarrow\,\Bustar\Km \), which can appear as a
  narrow peak at \(10\mevcc\ltsim Q\ltsim 70\mevcc\), shifted due to the \g
  undetected.

\begin{figure}[b]
\begin{center}
  \psfig{file=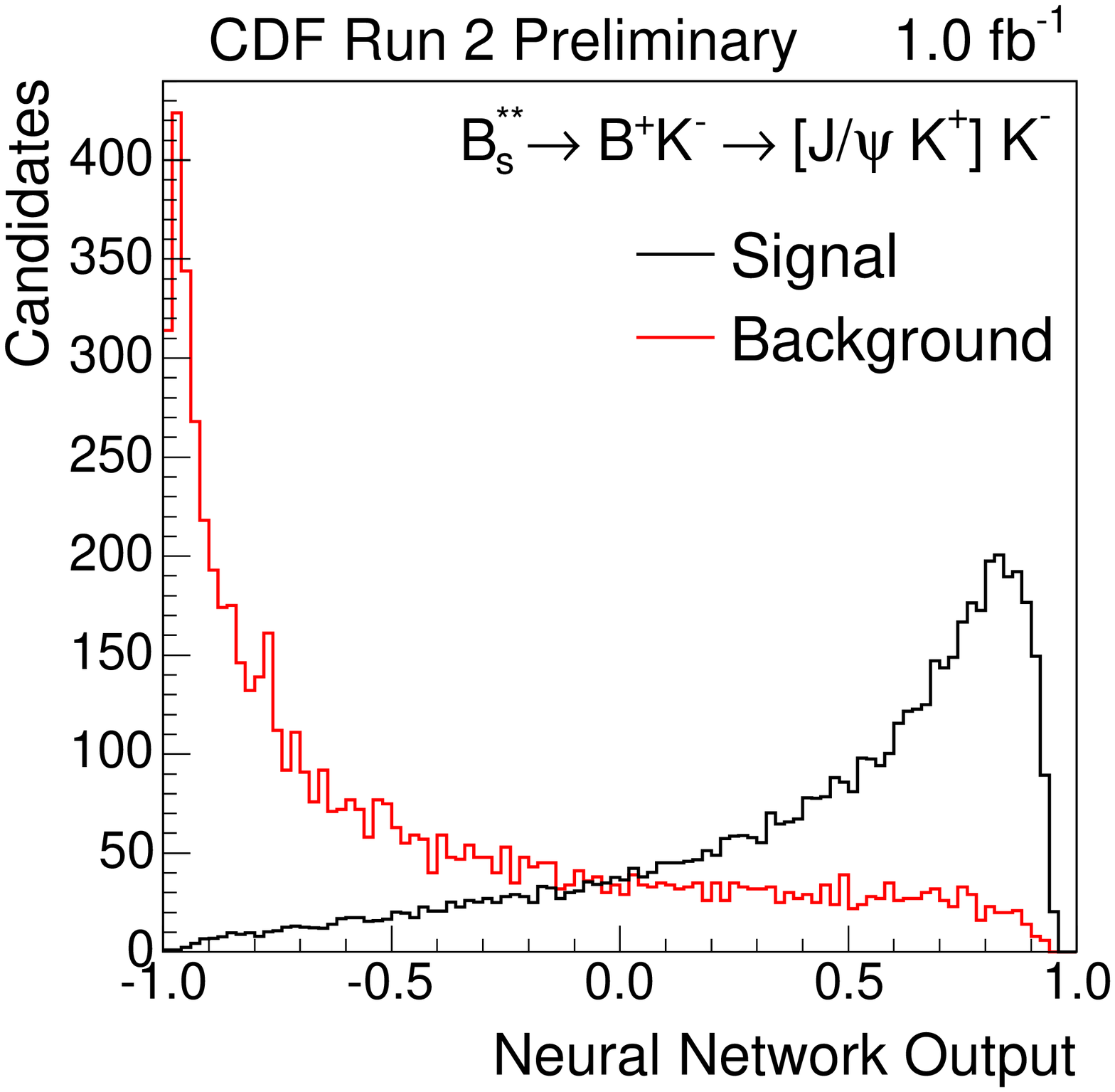,width=2.5in}
  \psfig{file=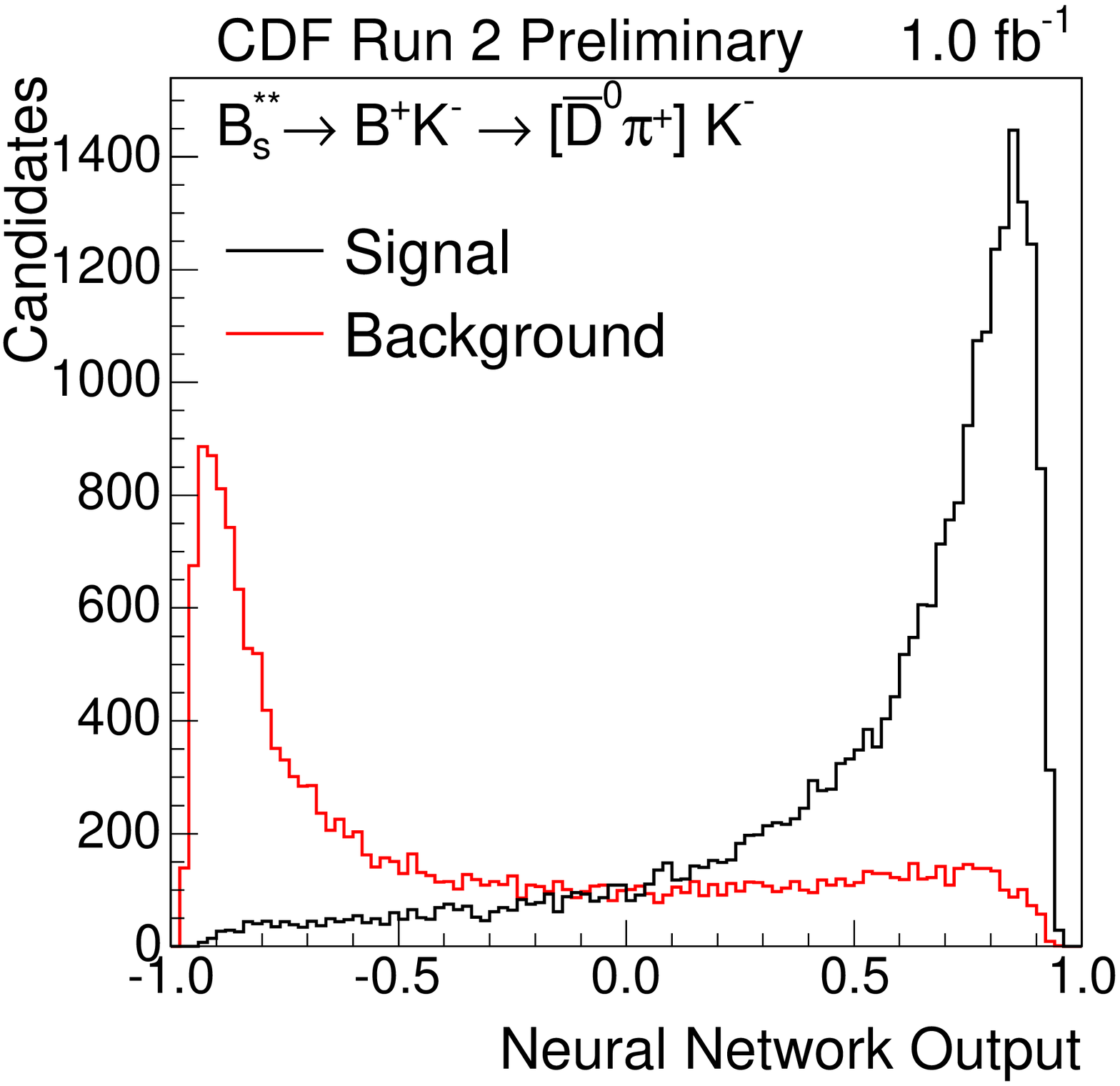,width=2.5in}
\end{center}
\caption{ Distribution of the neural network output for the signal and background events
          used for training of the neural network for the reconstructed decay modes
          \( \BsJ\rightarrow\,\Bu\Km,\,\Bu\rightarrow\jpsi\Kp \) and 
          \( \BsJ\rightarrow\,\Bu\Km,\,\Bu\rightarrow\Dzb\pip \).
        }
\label{fig:nn}
\end{figure}

  For the candidates \( B_{sJ}\rightarrow\,B^{(*)+}\Km \) the offline
  selections of reconstructed \( \Bu\Km \) pairs have been done
  separately for \( \Bu \) decays in the \( \Bu\rightarrow\jpsi\Kp \)
  or \( \Bu\rightarrow\,\Dzb\pip \) modes. In both cases the charm
  candidates and bottom $\Bu$ candidates have been subjected to
  3-dimensional vertex fits. 
  The candidates' topological quantities,
  kinematical variables, 
  and particle identification information have also
  been analyzed by neural networks built separately for the \Bu and 
  their parent \BsJ candidates~\cite{feindt:neurob}. The neural networks
  have been trained on background patterns using experimental data.
  To train the neural network for a signal pattern we have
  used Monte-Carlo simulated data having the same invariant mass distribution 
  as the experimental background events to avoid a bias to the signal. The distribution of the neural 
  network output for the neural networks trained individually for 
  two \BsJ decay modes is shown at Fig.~\ref{fig:nn}.
  For the final selection we apply the cuts on the neural network
  output \({\cal NN}\) and on the number of \BsJ candidates in the
  event. The cuts on the \({\cal NN}\) are chosen to maximize the
  figure of merit \( {S/\sqrt{S+B}} \) for candidates with 
  \( Q\in(60,70)\mevcc \) when the numerator is obtained 
  from Monte-Carlo and denominator is taken from data. 
  The cut values are found to be \( {\cal NN} >0.5 \) for 
  \( \BsJ\rightarrow\,\Bu\Km,\,\Bu\rightarrow\jpsi\Kp \) and 
  \( {\cal NN} >0.3 \) for 
  \( \BsJ\rightarrow\,\Bu\Km,\,\Bu\rightarrow\Dzb\pip \). 
  Only events with fewer than four \BsJ candidates 
  are allowed, in order to suppress further 
  the combinatorial background.
\section{Results}
\begin{figure}[b]
\begin{center}
  \psfig{file=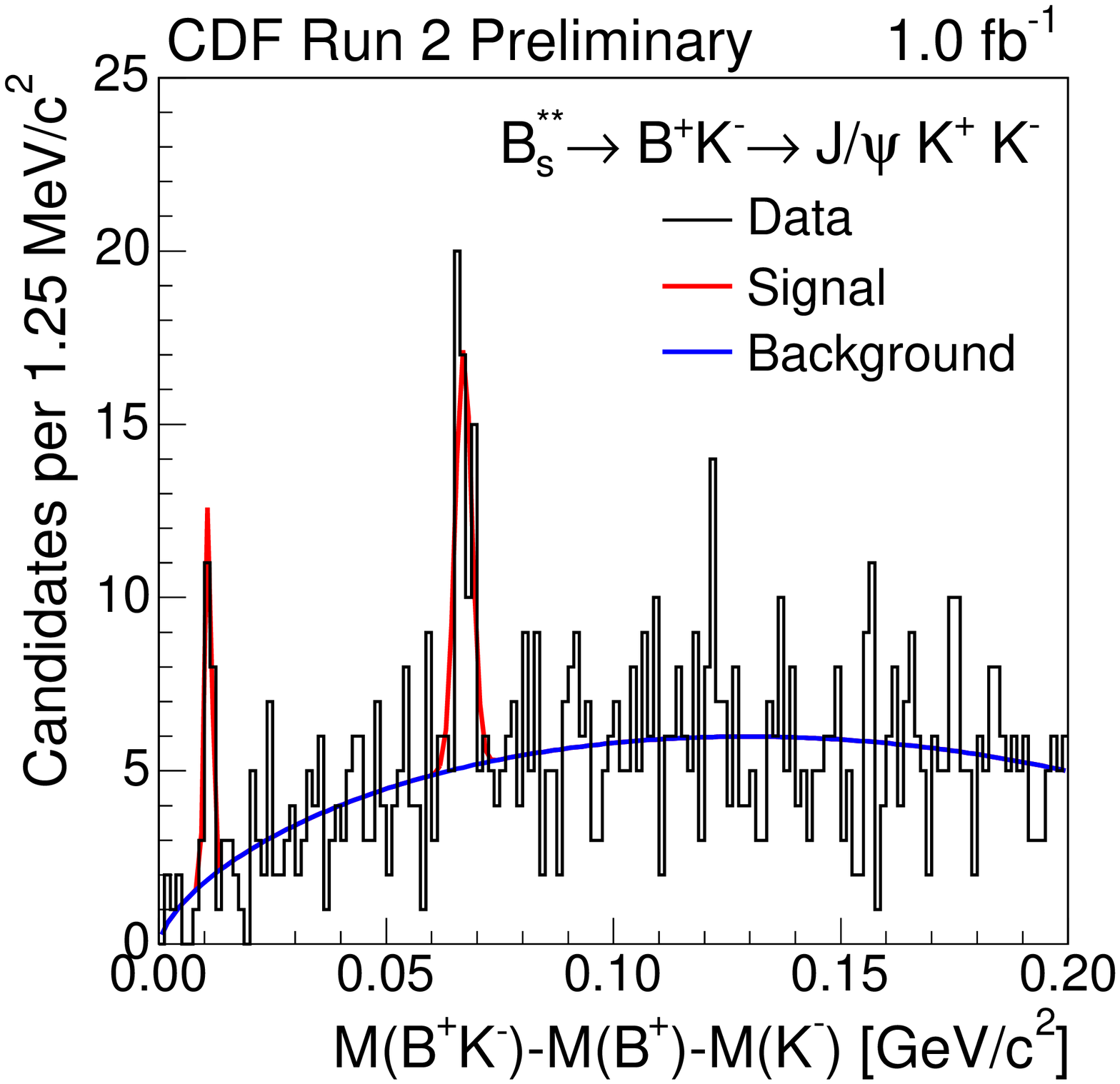,width=2.5in}
  \psfig{file=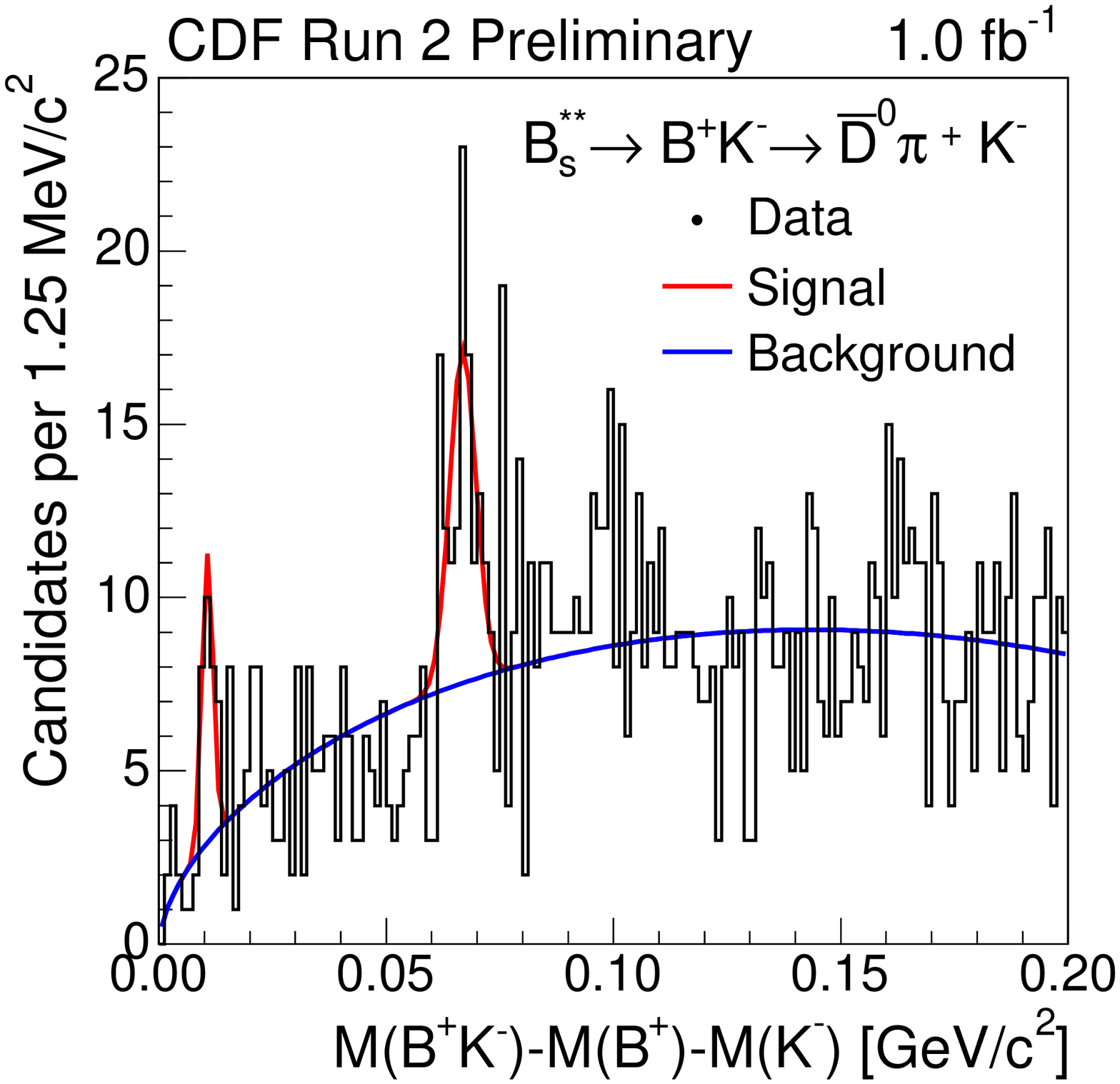,width=2.5in}
  \psfig{file=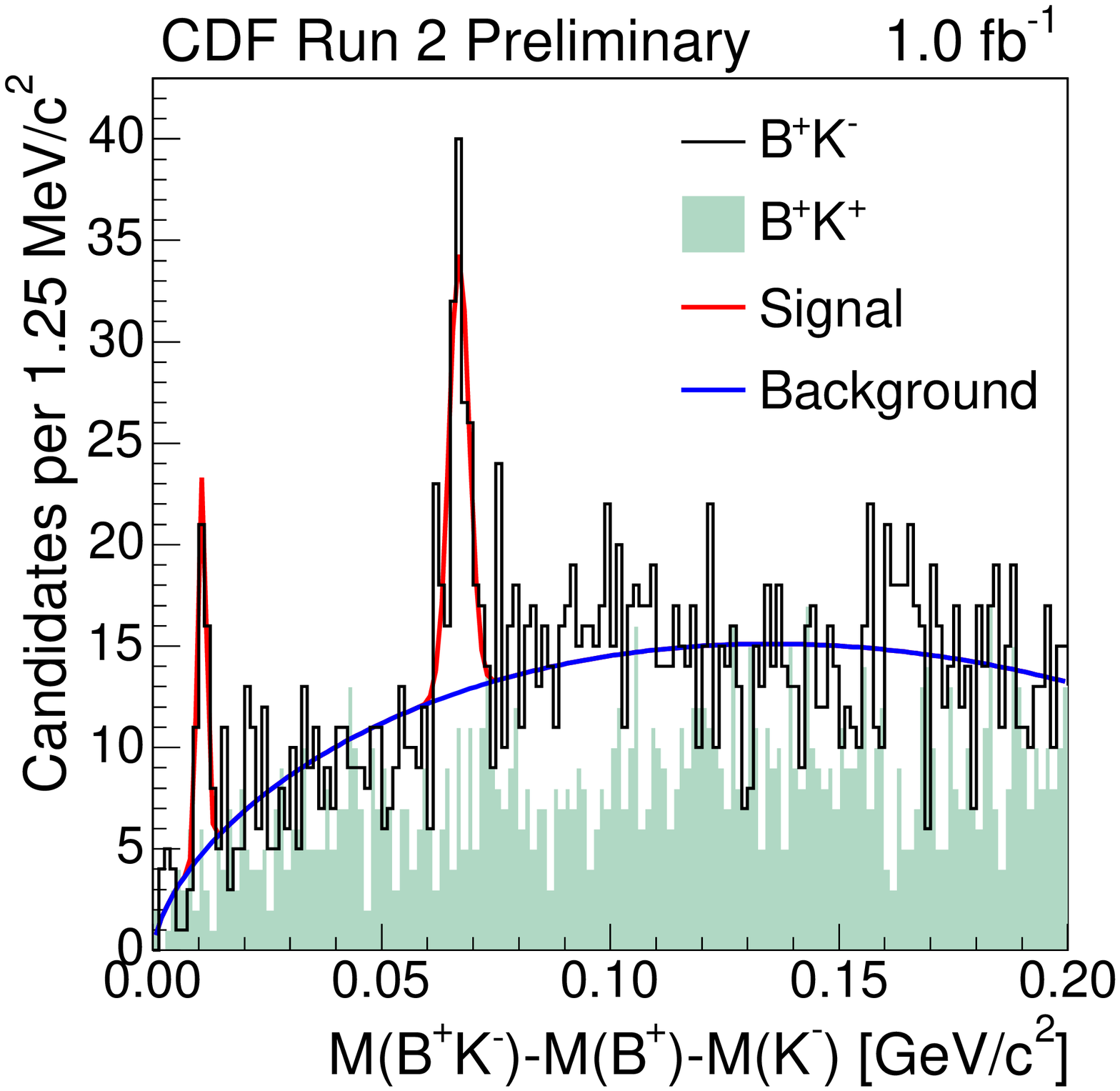,width=2.5in}
\end{center}
\caption{ The mass difference (Q-value) spectra for two modes 
          \( \BsJ\rightarrow\,\Bu\Km \), \( \Bu\rightarrow\jpsi\Kp\) or \( \Dzb\pip \)
          (two upper plots). The statistics of both modes is added and the final 
          result is shown at the bottom plot. The unbinned maximum 
          likelihood fit projections are superimposed on each plot.
        }
\label{fig:fits}
\end{figure}

  The $Q$-value spectra for every reconstructed decay mode of
  \( \BsJ\rightarrow\,\Bu\Km \) with 
  \( \Bu\rightarrow\jpsi\Kp\) or \(\Dzb\pip \) 
  are shown in Fig.~\ref{fig:fits} (two 
  upper plots).
  The spectra have been fit using an unbinned maximum likelihood method.
 \begin{table}
 \tbl{Summary of fit results, errors are statistical only \label{tab:fits}}
 {\begin{tabular}{@{}ccc@{}}
   \toprule
   State & \( \mathbf{\BsJone\rightarrow\,\Bustar\Km} \) & \( \mathbf{\BsJtwo\rightarrow\,\Bu\Km} \) \\ 
   \colrule
   $\Bu$~Mode & \multicolumn{2}{c}{\( \Bu\rightarrow\jpsi\Kp \)} \\ 
   \colrule
   \( Q [\mevcc]\) & \( 10.87\pm0.19 \) & \( 67.03\pm0.44 \) \\
   \( \sigma [\mevcc]\) & \( 0.64\pm0.25 \)  & \( 1.79\pm0.42 \) \\
   \( N_{evts} \) & \( 16.98\pm5.14 \) & \( 44.15\pm13.36 \) \\
   \colrule
   $\Bu$~Mode & \multicolumn{2}{c}{\( \Bu\rightarrow\Dzb\pip \)} \\ 
   \colrule
   \( Q [\mevcc]\) & \( 10.68\pm0.46 \) & \( 66.85\pm0.76 \) \\
   \( \sigma [\mevcc]\) & \( 1.18\pm0.56 \) & \( 2.88\pm0.75 \) \\
   \( N_{evts} \) & \( 20.66\pm7.12 \) & \( 55.74\pm19.20 \) \\
   \colrule
   Both $\Bu$~Modes & \multicolumn{2}{c}{ Combined statistics } \\ 
   \colrule
   \( Q [\mevcc]\) & \( \mathbf{10.73\pm0.21\stat} \) & \( \mathbf{66.96\pm0.39\stat} \) \\
   \( N_{evts} \) & \( \mathbf{36.4\pm9.0\stat} \) & \( \mathbf{94.8\pm23.4\stat} \) \\
   \botrule
 \end{tabular}}
 \end{table}
  As both decay modes have similar background shapes we add
  statistically both subsamples and proceed with the final fit, see
  the Fig.~\ref{fig:fits} bottom plot. The numerical results of the
  fits are summarized in Table~\ref{tab:fits}. Using a ratio of the
  original likelihood to the one of the fit with the null hypothesis, 
  \( -2\cdot\ln{ {\cal{L}}/{{\cal{L}}_{0}} } \), we obtain a significance
  of $6.3\sigma$ for the peak at $10.73\mevcc$ and of $7.7\sigma$ for the
  peak at $66.96\mevcc$. The statistical experiments with background 
  generated according to our data and the sole peak at $66.96\mevcc$
  yield a p-value of \( {2.13\cdot10^{-7}} \) for the newly established 
  signal at $10.73\mevcc$. 

\section{Summary}
  We have presented the observation of two narrow peaks in the mass
  difference \( M(\Bu\Km) - M(\Bu) - M(\Km) \) distribution.  The
  measured pattern of two peaks is interpreted as the signal of \(
  \BsJtwo\rightarrow\,\Bu\Km \) previously observed by other
  experiments and confirmed by our data and the signal of \(
  \BsJone\rightarrow\,\Bustar\Km \), reported here for the first
  time. Our measurements yielded:
  \begin{itemlist}
    \item \( {M(\BsJone)=5829.41\pm0.21\stat} \) 
          \( {\pm0.14\syst\pm0.6({PDG})\,\mevcc} \) 
    \item \( {M(\BsJtwo)=5839.64\pm0.39\stat} \)
          \( {\pm0.14\syst\pm0.5({PDG})\,\mevcc} \).
  \end{itemlist}
\section*{Acknowledgments}
  The author is grateful to Dr.~Michal~Kreps (Universit\"at Karlsruhe)
  for useful suggestions and comments made during the preparation of
  this talk. The author would like to thank Prof.~Sally~Seidel
  (Univ. of New Mexico) for support of this work and comments.
%


\end{document}